\documentclass{article}

\newcommand{\bt}{\begin{tabular}{c}}
\newcommand{\et}{\end{tabular}}

\newcommand{\np}{\newpage} 
 
\newcommand{\eb}{\ee\be } 
\newcommand{\ebp}{\rt.\ee\be\lt.} 
\newcommand{\bmat}{\lt ( \begin{array} }
\newcommand{\emat}{  \end{array} \rt )}

\newcommand{\ED}{\end{document}}

\newcommand{\oy}{{\ov \y}}

\newcommand{\A}{{\ov A}}

\renewcommand{\a}{\alpha}	
\renewcommand{\b}{\beta}
\newcommand{\g}{\gamma}
\renewcommand{\d}{\delta}
\newcommand{\e}{\epsilon}
\newcommand{\ve}{\varepsilon}

\newcommand{\q}{\theta}

\renewcommand{\r}{\rho}

\newcommand{\f}{\phi}

\newcommand{\y}{\psi}
\newcommand{\w}{\omega}
\newcommand{\G}{\Gamma}
\newcommand{\D}{\Delta}

\renewcommand{\P}{\Pi}

\newcommand{\la}{\label}
\newcommand{\ci}{\cite}

\newcommand{\ds}{\documentstyle}	
\newcommand{\fr}{\frac}

\newcommand{\pa}{\partial}
\newcommand{\ov}{\overline}
\newcommand{\be}{\begin{equation}}
\newcommand{\ee}{\end{equation}}
\newcommand{\ba}{\begin{array}} 
\newcommand{\ea}{\end{array}}
\newcommand{\bea}{\begin{eqnarray}}
\newcommand{\eea}{\end{eqnarray}}
\newcommand{\ra}{\rightarrow}
\newcommand{\Ra}{\Rightarrow}

\newcommand{\Lra}{\Leftrightarrow}
\newcommand{\lt}{\left}
\newcommand{\rt}{\right}

\newcommand{\ben}{\begin{enumerate}}
\newcommand{\een}{\end{enumerate}}
\newcommand{\bitem}{\begin{itemize}}
\newcommand{\eitem}{\end{itemize}}
\newcommand{\articlenumber}{\\ }

\newcommand{\articletitle}{ A supersymmetric version of the quark model, and supersymmetry breaking for the Leptons, Baryons and Hadronic Mesons: Cybersusy V}

\begin{document}
\makeatletter	   
\renewcommand{\ps@plain}{%
\renewcommand{\@oddhead}{{\articlenumber  \hspace{1cm} }\hspace{1cm}  \hfil\textrm{\thepage}} 
\renewcommand{\@evenhead}{\@oddhead}
\renewcommand{\@oddfoot}{\textrm{\articlenumber \hspace{1cm}  }\hspace{1cm} \hfil\textrm{\thepage}}
\renewcommand{\@evenfoot}{\@oddfoot}}
\makeatother    
\title{\articletitle \articlenumber}
\author{ J. A.  Dixon\footnote{jadix@telus.net}\\ Dixon Law Firm\footnote{Fax: (403) 266-1487} \\1020 Canadian Centre\\
833 - 4th Ave. S. W. \\ Calgary, Alberta \\ Canada T2P 3T5 }
\maketitle
\pagestyle{plain}
\Large
 
\abstract{\large  Cybersusy is a new mechanism for supersymmetry breaking in the standard supersymmetric model (SSM).  Here we note that the superpotential for the SSM has a set of thirteen invariances, five of which are well known, and eight of which are new. The  eight new invariances generate a sort of supersymmetric quark and lepton model, together with supersymmetry breaking that makes the squarks and sleptons very heavy.  This breaking regenerates the standard model out of the supersymmetric standard model, except that the gauge particles are not yet included in this reduction. In this paper, it is shown that, with some continued effort, cybersusy will make some predictions for baryon masses that may actually be wrong, so that it is a supersymmetry breaking theory that can be proved wrong!  This is the fifth paper in what was intended to be a series of four papers on cybersusy.}

\Large
\section{Why does the Standard Model have such a bizarre  particle content?}
\la{fewfwfwefwefkthklhrtkl}
It is notorious that the particle content of the standard model is rather strange, and little explanation or motivation for it has been available.  With a huge amount of effort, from both theorists and experimentalists, its detailed structure has been determined and parametrized in terms of quantum field theory 
\ci{physletreview}.  

The supersymmetric standard model (SSM) is just a repetition of this strangeness, except that it adds a whole set of unobserved superpartners, which takes the standard model from strange to absurd. 

 But supersymmetry is a very tempting generalization of the standard model, because of the superstring, and because it is such an interesting symmetry in its own right.  Since we are very comfortable with the idea of a broken symmetry, it is natural to guess that a suitable kind of supersymmetry breaking will cure the absurdity of the SSM.

But it is clear that this cure is not the spontaneous breaking of supersymmetry, because that entails a huge vacuum energy, whereas supersymmetry without spontaneous breaking has zero vacuum energy, which is very near the truth.  There are also embarrassing mass sum rules that arise from  the spontaneous breaking of supersymmetry.  These  make it very hard to make the theory consistent with experiment, even if one chooses to ignore the vacuum energy problem.

It is also clear that there is a great deal left to be desired from the other popular method for breaking supersymmetry, which is the addition of soft supersymmetry breaking masses to the squarks and the sleptons in the SSM.  These explicitly break supersymmetry, but really the only motivation for them is the desire to break supersymmetry.  They seem to have no very satisfactory origin in the theory\footnote{\large However, from one point of view, cybersusy appears to generate something very much like this.  As is shown below, cybersusy generates an effective theory with a supersymmetric quark model for hadrons, with supersymmetry breaking.  Since the mechanism also applies to the quarks, it looks a lot like soft breaking, except that the particle content is more complicated for the quark multiplets in cybersusy.  Also, there is another breaking for composite hadrons that probably gives a different, more reliable, result. This is discussed in section 
\ref{erjiopejioperge} below}.

It is also notorious that many current  models of particle theory (e.g. superstring theories, explicitly broken supersymmetry) are very difficult to prove wrong, because they do not make sufficiently well-defined predictions.

Cybersusy is a new mechanism for breaking supersymmetry.  It has its origin in the BRS cohomology of supersymmetry, as applied to  composite operators.  The breaking is extracted from the algebra of the composite operators by using the technique of effective actions to analyze new symmetries. The composite operators that arise from the BRS cohomology do in fact closely resemble the physical particles that we see from the standard model, in a supersymmetric form. The BRS cohomology of a theory naturally gives rise to the physically relevant degrees of freedom of that theory. This is obvious, in a sense, because such physically relevant operators should be supersymmetric, and also not vanish by the equations of motion, and that is what the BRS cohomology finds.   Part of the new material in this paper is to demonstrate this fact for the leptons and for the hadrons (both mesons and baryons). 

The cybersusy mechanism is more closely related to explicit breaking than it is to spontaneous breaking, and so it does not give rise to a vacuum energy.  However the breaking does not simply take the form of changing the masses of the sleptons and squarks.  The breaking arises from transformations which are very much like a set of supersymmetry anomalies.  
These break rigid or global supersymmetry, and they arise for the special case of the SSM. A dichotomy  arises, which  is very much like an anomaly.  One cannot have massive dotspinor supermultiplets, describing the bound states of the SSM, at the same time that the theory  also contains the spontaneous breaking of gauge symmetry.

A bonus of cybersusy, and the BRS cohomology, is that they
seem to supply some motivation for the strange structure of the standard model, as will be seen below. It appears that the SSM is `rigged' to produce supersymmetry breaking through cybersusy. 

 A further bonus of cybersusy is that {\bf it can be proved wrong} (or, possibly, not wrong, yet) with relatively little further effort,  as is set out in the conclusion to this paper.

\section{Introduction}

In the first four papers on cybersusy \ci{cybersusyI}\ci{cybersusyII}\ci{cybersusyIII}\ci{cybersusyIV}, it was shown that there is a scheme of supersymmetry breaking that is based on the cybersusy algebra. This was illustrated with a calculation of the mass spectrum for the three electron flavours and the three neutrino flavours.  That algebra is closely linked to the standard supersymmetric model (SSM) and its remarkable superpotential. The mass spectrum is quite simple when one assumes that the relevant mixing matrices are diagonal.  

 Brief remarks were also made about  the baryons, but no attempt was made to determine the spectrum there. It was, however, argued that the supersymmetry breaking will be in the baryon spectrum, not in the elementary spectrum of the quarks and squarks.

In this paper we discuss the superpotential in this context.  There are a number of simple things that can be said here, and we can make some important progress in an easy way.  This is a revised version of this paper, and it has some important changes from the original \ci{cybersusyV}.

We will start this paper with a discussion of some interesting properties of the superpotential of the SSM, and then return to a discussion of  cybersusy.

Here is the superpotential for the SSM:
\[
P_{\rm SSM}   =
g \e_{ij} H^i K^j J
- g_{\rm J} m^2 J
+
p_{p q} \e_{ij} L^{p i} H^j P^{ q} 
\]\be
+
r_{p q} \e_{ij} L^{p i} K^j R^{ q}
+
t_{p  q} \e_{ij} Q^{c p i} K^j T_c^{ q}
+
b_{p  q} \e_{ij} Q^{c p i} H^j B_c^{ q}
\la{suppotential}
\ee
and here is a table of the quantum numbers:
\be
\begin{tabular}{|c|c|c
|c|c|c
|c|c|c|}
\hline
\multicolumn{8}
{|c|}{Table of the Chiral Superfields in the SSM
}
\\
\hline
\multicolumn{8}
{|c|}{ \bf Superstandard Model, Left ${\cal L}$
Fields}
\\
\hline
{\rm Field} & Y 
& {\rm SU(3)} 
& {\rm SU(2)} 
& {\rm F} 
& {\rm B} 
& {\rm L} 
& {\rm D} 
\\
\hline
$ L^{pi} $& -1 
& 1 & 2 
& 3
& 0
& 1
& 1
\\
\hline
$ Q^{cpi} $ & $\fr{1}{3}$ 
& 
3 &
2 &
3 &
 $\fr{1}{3}$
& 0
& 1
\\\hline
$J$
& 0 
& 1
& 1
& 1
& 0
& 0
& 1
\\
\hline
\multicolumn{8}
{|c|}{ \bf Superstandard Model, Right 
${\cal R}$
Fields}
\\
\hline
$P^{ p}$ & 2 
& 1
& 1
& 3
& 0
& -1
& 1
\\
\hline
$R^{ p}$ & 0 
& 1
& 1
& 3
& 0
& -1
& 1
\\

\hline
$T_c^{ p}$ & $-\fr{4}{3}$ 
& ${\ov 3}$ &
1 &
3 &
 $-\fr{1}{3}$
& 0
& 1
\\
\hline
$B_c^{ p}$ & $\fr{2}{3}$ 
& ${\ov 3}$ 
&
1 &
3 &
 $- \fr{1}{3}$
& 0
& 1
\\
\hline
$H^i$ 
& -1 
& 1
& 2
& 1
& 0
& 0
& 1
\\
\hline
$K^i$ 
& 1 
& 1
& 2
& 1
& 0
& 0
& 1
\\
\hline
\end{tabular}
\\
\la{qefewfqwefwrrr}
\ee

There are several aspects to a kind of emerging logic behind the particle content of the supersymmetric standard model.  First we note that:
\ben
\item
The Higgs fields $H^i$ and $K^i$ are needed to give masses to the leptons and quarks, and they are also needed to spontaneously break 
$SU(3) \times SU(2) \times U(1)$ down to $SU(3) \times   U(1)$.
\item
The standard model has left/right asymmetry.  This is repeated twice:\ben
\item
  The quarks are in left weak $SU(2)$ doublets  $Q^{c p i} $ and right weak $SU(2)$ singlets $T_c^{ q}$ and $B_c^{ q}$.
\item
The leptons are in left weak $SU(2)$ doublets  $L^{ p i} $ and right weak $SU(2)$ singlets $P^{ q}$ and $R^{ q}$ (we need to assume that the neutrinos have Dirac type masses). 
\een
\item
The Higgs field $J$ is needed  to  trigger the 
spontaneous breakdown of  
$SU(3) \times SU(2) \times U(1)$ down to $SU(3) \times   U(1)$. The spontaneous breakdown requires that the symmetry breaking term $- g_{\rm J} m^2 J
$ be present in the superpotential of the SSM.
\een

When one looks at the BRS cohomology of the theory, one also finds that:
\ben
\item
The two Higgs $SU(2)$ doublets  $H^i, K^i$  and the Higgs singlet $J$, and the left/right asymmetry of the quarks and leptons,  allow the construction of the eight cybersusy generators in section \ref{adfafdsfljjklasdf} below.
\item
The Higgs field $J$ is also an essential ingredient in supersymmetry breaking in cybersusy, since only generators that include $\fr{\pa}{\pa J}$ give rise to the cybersusy algebra.  The spontaneous breakdown of gauge symmetry also results in an explicit breaking of supersymmetry as discussed in the four papers \ci{cybersusyI}\ci{cybersusyII}\ci{cybersusyIII}\ci{cybersusyIV}. This breaking is not a spontaneous breaking of supersymmetry, and so it does not result in a vacuum energy problem.
\een

In other words,   the left/right asymmetry, and the Higgs fields, are needed to generate supersymmetry breaking for the leptons and the hadrons, in a way that gives rise to a universe that does not suffer from a huge cosmological constant.

\section{The  Lie Algebra of invariance for the superpotential for the supersymmetric  standard model}
\la{qegrtuyjhki}

It was shown in \ci{cybersusyII} that the generators of the simplest non-derivative sector of the  
BRS cohomology, for any supersymmetric theory of massless, chiral, scalar superfields,  are isomorphic to the generators of the Lie algebra of invariance for the superpotential of that theory.

We can quickly repeat the main pieces of the argument here for clarity.  It was shown in \ci{cybersusyII}  that, for a general theory, there are chiral dotted spinor pseudosuperfields ${\hat \f}_{i \dot \a}$ for each spinor in the theory.  They transform as
\be
\d {\hat \f}_{i \dot \a} = \d_{\rm SS} {\hat \f}_{i \dot \a}
-g_{ijk} {\hat A}^{j}{\hat A}^{k} \ov C_{\dot \a}
\la{fwewefwew}
\ee
where $\d_{\rm SS} $ is the usual superfield transformation for a dotted chiral spinor superfield.

As explained in \ci{cybersusyII}, the expansion  of ${\hat \f}_{i \dot \a}$ in terms of the superspace parameters $\q$ is:
\be
{\hat \f}_{i \dot \a}(x) = \ov \y_{\dot \a i}(y) +
\q^{\b} 
\lt (
\pa_{\b \dot \a} \A_i(y) 
+ \ov C_{\dot \a} Y_{i  \b}(y) \rt )
- \fr{1}{2} \q^{\g} \q_{\g} 
\G_{i}(y)  
\ov C_{\dot \a} 
\la{weird}
\ee
 So an expression of the form
\be
{\hat \w}_{\dot \a} = f^{i}_{j} {\hat A}^j {\hat \f}_{i \dot \a}
\ee
will transform as a chiral dotted spinor superfield, with the usual transformation $\d_{\rm SS} $, without the extra terms in  
(\ref{fwewefwew}), 
if the following constraint
is satisfied:
\be
f^{i}_{j} {\hat A}^j  g_{ikl} {\hat A}^{k}{\hat A}^{l}=0
\la{fwefwefwefwewe}
\ee
We can drop the hats and just use the lowest weight fields, and write this in the equivalent form (which  arises from the spectral sequence):
\be
d_3 {  \w}_{\dot \a}
= d_3f^{i}_{j} A^j  \ov \y_{i \dot \a}
=0
\la{wfweffgwfw}
\ee
where
\be
d_3 =  \ov C_{\dot \a} g_{ikl} A^{k}A^{l}\ov \y_{i \dot \a}^{\dag}
\ee

The term $g_{ijk} A^{j}A^{k} $ is just the derivative of the superpotential
\be
P_{\rm SP}= g_{ijk} {  A}^{i}{  A}^{j}{  A}^{k}
\ee
 with respect to a field $A^i$.  So equation 
(\ref{fwefwefwefwewe}) can be written as a requirment that the superpotential satisfy the invariance equation:
\be
{\cal L}_{f} P_{\rm SP}=0
\ee
where ${\cal L}_{f}$ is a generator of the Lie algebra of invariances of the superpotential, and it has the form:
\be
{\cal L}_{f} =
f^{i}_{j} A^j   \fr{\pa}{\pa A^i} 
\ee

So now we shall look at the invariances of the superpotential for the SSM.  Throughout this paper, the complex conjugates of the statements are also true, and we shall not write the complex conjugates out in detail.  

\subsection{Ordinary Symmetry Generators for the superpotential of the SSM}

\la{hwhuyhjytjtjy}

As is well known, the superpotential of the standard model is chosen to be invariant under a number of symmetries.   It turns out that there are two general kinds of generators for the  Lie Algebra of invariance of the superpotential for the standard supersymmetric model.  First we will discuss   the well-known invariances of the SSM.

A glance at the Table in Equation (\ref{qefewfqwefwrrr}) reminds us that there are some obvious generators for invariance of the superpotential for the standard supersymmetric model.
These are the generators   of the groups $SU(3) \times SU(2) \times U(1)$ and counting operators for lepton number and baryon number.  
Here they are:
\ben
\item  Strong SU(3):
\be
 {\cal C}^A = T^{A c}_{d}\lt (
  Q^{dpi} \fr{\pa}{\pa Q^{cpi}}
+  B^{p}_{c} \fr{\pa}{\pa  B^{p}_{d} }
+   T^{p}_{c} \fr{\pa}{\pa T^{p}_{d}}\rt )
\ee
where $T^{A c}_{d}$ are the hermitian $3 \times 3$ matrix generators of the Lie Algebra of SU(3),
\item  Weak SU(2)
\be
 {\cal I}^a = t^{ai}_{j}\lt (
H^j \fr{\pa}{\pa H^i}
+K^j \fr{\pa}{\pa K^i}
+L^{pj} \fr{\pa}{\pa L^{pi}}
+Q^{cpj} \fr{\pa}{\pa Q^{cpi}}
\rt )
\la{calia2}
\ee
where $t^{ai}_{j}$ are the hermitian $2 \times 2$ matrix generators of the Lie Algebra of SU(2),
\item  Hypercharge U(1)
\be
 {\cal Y} = \lt (
- H^i \fr{\pa}{\pa H^i}
+ K^i \fr{\pa}{\pa K^i}
-L^{pi} \fr{\pa}{\pa L^{pi}}
\ebp
+ 2 P^{p} \fr{\pa}{\pa P^{p}}
+ \fr{1}{3} Q^{cpi} \fr{\pa}{\pa Q^{cpi}}
+ \fr{2}{3} B^{p}_{c} \fr{\pa}{\pa  B^{p}_{c} }
- \fr{4}{3} T^{p}_{c} \fr{\pa}{\pa T^{p}_{c}}
\rt )
\ee
\item Lepton Number
\be
 {\cal L}= \lt (
L^{pi} \fr{\pa}{\pa L^{pi}}
- P^{p} \fr{\pa}{\pa P^{p}}
- R^{p} \fr{\pa}{\pa R^{p}}
\rt )
\ee
\item Baryon Number
\be 
  {\cal B}=
 \fr{1}{3} Q^{cpi} \fr{\pa}{\pa Q^{cpi}}
- \fr{1}{3} B^{p}_{c} \fr{\pa}{\pa  B^{p}_{c} }
- \fr{1}{3} T^{p}_{c} \fr{\pa}{\pa T^{p}_{c}}
\ee
\een
It is easy to verify that  these satisfy the relations:
\be
{\cal I}^a 
P_{\rm SSM}={\cal C}^A 
P_{\rm SSM}={\cal Y} 
P_{\rm SSM}={\cal B} 
P_{\rm SSM}={\cal L} 
P_{\rm SSM}=0
\ee
Note that we do {\bf not} need to leave  the symmetry breaking term $- g_{\rm J} m^2 J
$ out for these calculations.  Also note that none of the above generators have a term of the form $\fr{\pa}{\pa J}$.  This is a consequence of the fact that $J$ has zero lepton number and zero baryon number, and also that $J$  is invariant under the groups    $SU(3) \times SU(2) \times U(1)$.

\subsection{Left/right assymetric  generators of the Lie algebra for the invariance of the superpotential of the SSM: quark and lepton invariants}
\la{adfafdsfljjklasdf}

Now we shall discuss some  more subtle invariances of the SSM.
Because of its form, we can quickly find eight more independent generators of invariance of the SSM superpotential, and these have a one-to-one map onto the leptons and quarks.

There are two `left handed' quark generators and two `right handed' quark generators. There are two of each because the quarks can be up or down under weak SU(2).  So in a sense the generators split the left handed doublets into the charge eigenstates.  

Similarly, there are two `left handed' lepton generators and two `right handed' lepton generators. There are two of each because the leptons can be up or down under weak SU(2).  Again, in a sense, the generators split the left handed doublets into the charge eigenstates for the leptons.  

As will be seen in detail below, the BRS cohomology is performing the task of generating a kind of supersymmetric quark (and lepton) model through this mechanism.

But to discuss this we must first introduce the generators.

\subsubsection{`Right handed' Lepton Generators}

The first basic generator is the Positron Generator ${\cal L}_{P,s}$:
\be
{\cal L}_{P,s}=p_{sq} P^q
\fr{\pa}{\pa J} + 
g   K^j \fr{\pa}{\pa L^{sj}} 
\la{fqewfwewefwopjpjopj}
\ee
Let us demonstrate that this is indeed a generator of the invariance of the superpotential (\ref{suppotential}).
Note that we must leave the symmetry breaking term $- g_{\rm J} m^2 J
$ out for this calculation.\[
{\cal L}_{P,s} P_{\rm SSM} 
\]
\[
=\lt \{
p_{sq} P^q
\fr{\pa}{\pa J} + 
g   K^k \fr{\pa}{\pa L^{sk}} 
\rt \}
\lt \{ g \e_{ij} H^i K^j J
+
p_{p q} \e_{ij} L^{p i} H^j P^{ q} 
\rt.\]\be\lt.
+
r_{p q} \e_{ij} L^{p i} K^j R^{ q}
+
t_{p  q} \e_{ij} Q^{c p i} K^j T_c^{ q}
+
b_{p  q} \e_{ij} Q^{c p i} H^j B_c^{ q}
\rt \}
\ee
and this reduces immediately to
\[
{\cal L}_{P,s} P_{\rm SSM} 
\]
\[
=
p_{sq} P^q
\fr{\pa}{\pa J} g \e_{ij} H^i K^j J
\]
\be
+ 
g   K^k \fr{\pa}{\pa L^{sk}}  \lt \{  
p_{p q} \e_{ij} L^{p i} H^j P^{ q} 
+
r_{p q} \e_{ij} L^{p i} K^j R^{ q}
\rt \}
\ee
\[
=
p_{sq} P^q  g \e_{ij} H^i K^j  
\]
\be
+ 
g   K^i
p_{s q} \e_{ij}   H^j P^{ q} 
+ 
g   K^i
r_{s q} \e_{ij}   K^j R^{ q}
=0
\ee

There are seven more generators like 
(\ref{fqewfwewefwopjpjopj}), and their invariance works in a similar way. We shall list them all here. The other right leptonic generator is the Right Neutrino Generator ${\cal L}_{R,s}$:
\be
{\cal L}_{R,s}=r_{sq} R^q
\fr{\pa}{\pa J} + 
g   H^j \fr{\pa}{\pa L^{sj}} 
\ee

\subsubsection{`Left handed' Lepton Generators}

 The first basic `left handed' lepton generator generator is the
Electron Generator ${\cal L}_{E,s}$:

\be
{\cal L}_{E,s}= p_{qs} \e_{ij} H^i L^{qj} 
\fr{\pa}{\pa J} + 
g \e_{ij} H^i K^j \fr{\pa}{\pa P^{s}} 
\ee
Now let us demonstrate that this is also indeed a generator of the invariance of the superpotential (\ref{suppotential}). Note that we must again leave the symmetry breaking term $- g_{\rm J} m^2 J
$ out for this calculation.
\[
{\cal L}_{E,s}
 P_{\rm SSM} 
\]
\[
=\lt \{
 p_{qs} \e_{ij} H^i L^{qj} 
\fr{\pa}{\pa J} + 
g \e_{ij} H^i K^j \fr{\pa}{\pa P^{s}} 
\rt \}
\lt \{ g \e_{ij} H^i K^j J
+
p_{p q} \e_{ij} L^{p i} H^j P^{ q} 
\rt.\]\be\lt.
+
r_{p q} \e_{ij} L^{p i} K^j R^{ q}
+
t_{p  q} \e_{ij} Q^{c p i} K^j T_c^{ q}
+
b_{p  q} \e_{ij} Q^{c p i} H^j B_c^{ q}
\rt \}
\ee
and this reduces immediately to
\[
{\cal L}_{E,s} P_{\rm SSM} 
\]
\[
=
 p_{qs} \e_{kl} H^k L^{ql} 
\fr{\pa}{\pa J} 
\lt (
g \e_{ij} H^i K^j J
\rt )
\]
\be 
+ 
g \e_{kl} H^k K^l \fr{\pa}{\pa P^{s}} 
\lt (
p_{p q} \e_{ij} L^{p i} H^j P^{ q} 
\rt )
=0
\ee
Then there is the left handed neutrino generator:
\be
{\cal L}_{N,s}= r_{qs} \e_{ij} K^i L^{qj} 
\fr{\pa}{\pa J} + 
g \e_{ij} H^i K^j \fr{\pa}{\pa R^{s}} 
\ee

\subsubsection{`Right handed' Quark Generators}

For right handed quarks, we have the Bottom Generator ${\cal L}_{B,sc}$:
\be
{\cal L}_{B,sc}
=b_{sq} B^q_c
\fr{\pa}{\pa J} + 
g   K^j \fr{\pa}{\pa Q^{scj}} 
\ee
Let us check that
\[
{\cal L}_{B,sc}P_{\rm SSM} 
=\lt \{
b_{sq} B^q_c
\fr{\pa}{\pa J} + 
g   K^j \fr{\pa}{\pa Q^{scj}} 
\rt\}
\]\[
\lt \{ g \e_{ij} H^i K^j J
+
p_{p q} \e_{ij} L^{p i} H^j P^{ q} 
\rt.\]
\be\lt.
+
r_{p q} \e_{ij} L^{p i} K^j R^{ q}
+
t_{p  q} \e_{ij} Q^{c p i} K^j T_c^{ q}
+
b_{p  q} \e_{ij} Q^{c p i} H^j B_c^{ q}
\rt \}
=0
\ee

This is
\[
{\cal L}_{B,sc}P_{\rm SSM} 
= 
b_{sq} B^q_c
 g \e_{ij} H^i K^j 
\]
\be
+ 
g   K^j \fr{\pa}{\pa Q^{scj}} 
\lt \{ 
t_{p  q} \e_{ij} Q^{c p i} K^j T_c^{ q}
+
b_{p  q} \e_{ij} Q^{c p i} H^j B_c^{ q}
\rt \}
\ee
which reduces to
\be
{\cal L}_{B,sc}P_{\rm SSM} 
\eb
= 
b_{sq} B^q_c
 g \e_{ij} H^i K^j 
+ 
g   K^i  
t_{s  q} \e_{ij}  K^j T_c^{ q}
+ 
g   K^i  
b_{s  q} \e_{ij}  H^j B_c^{ q}
=0
\ee

Then there is also  the  Top Generator ${\cal L}_{T,sc}$:
\be
{\cal L}_{T,sc}
=t_{sq} T^q_c
\fr{\pa}{\pa J} + 
g   H^j \fr{\pa}{\pa Q^{scj}} 
\ee

\subsubsection{`Left handed' Quark Generators}

For left  handed quarks, we have the Up Generator
 ${\cal L}_{U,s}^c$:
\be
{\cal L}^c_{U,s}= t_{qs} \e_{ij} K^i Q^{cqj} 
\fr{\pa}{\pa J} + 
g \e_{ij} H^i K^j \fr{\pa}{\pa T^{s}_c} 
\ee
and the  Down Quark Generator ${\cal L}_{D,s}$
\be
{\cal L}^c_{D,s}= b_{qs} \e_{ij} H^i Q^{cqj} 
\fr{\pa}{\pa J} + 
g \e_{ij} H^i K^j \fr{\pa}{\pa B^{s}_c} 
\ee

\subsection{Generation of the whole Lie Algebra}

 It seems a reasonable conjecture that the above are all the basic generators of the Lie algebra of the invariance of the Superpotential for the SSM. There are five `ordinary' generators in subsection \ref{hwhuyhjytjtjy}, and eight more `J-invariance' generators in subsection
 \ref{adfafdsfljjklasdf}.  Let us label them ${\cal L}_{n}$ where $n=1,2,\cdots 13$.  It is easy to verify that taking 
 the commutator of any two basic generators
 ${\cal L}_{n}$  generates nothing different from the original 13 ${\cal L}_{n}$.  
\la{gqreghhgkprthhrt}

\section{Reflections on the Standard Model}

What is special about the SSM that allows these `J-invariances' which are ismorphic to the quarks and leptons? It would be interesting to see how many changes could be made while retaining this isomorphism.  In the SSM, a large role is being played by the weak group SU(2) and its invariant tensor $\e_{ij}$.  Having the ability to get masses from the two Higgs fields $H^i$ and $K^i$ comes from having left doublets and right singlets for the quarks and leptons.  For example, if both left and right quarks and left and right leptons were doublets, masses would require a field like $H^{(ij)}$  or $H$ to couple to both left and right, and then it not easy to see how to get these `J-invariances', together with other necessary or desirable properties.  A complete discussion of this would be tricky, of course, because there are so many possibilities, and it is not clear, and it is not likely to become clear,  what properties are necessary, or desirable, when designing a standard model.

\section{Dotspinors and Cybersusy}

As was noted in \ci{cybersusyII}, and discussed briefly above in section \ref{qegrtuyjhki}, the existence of a solution for the cohomology constraints for the simple generators is equivalent to the existence of a generator for the  Lie Algebra of the invariance of the superpotential for a given massless chiral theory. 

Here is how this works. Suppose we have a generator $ {\cal I}^a$ as in  (\ref{calia2}),  or ${\cal L}_{P,s} $ as in (\ref{fqewfwewefwopjpjopj}),
which satisfies the invariance equation:
\be
{\cal I}^a P_{\rm SSM}   =0
\ee
\be
{\cal L}_{P,s}P_{{\rm SSM}}   =0
\ee
Let us take the simpler example:
\be
{\cal L}_{P,s}=p_{sq} P^q
\fr{\pa}{\pa J} + 
g   K^j \fr{\pa}{\pa L^{sj}} 
\ee
Then we can construct a chiral dotted spinor superfield from it in the following way:
\ben
\item
For each derivative in ${\cal L}_{P,s}$, make a replacement 
as follows
\be
\fr{\pa}{\pa J} \Rightarrow \oy_{J \dot \a}
\ee
\be
\fr{\pa}{\pa L^{sj}}  \Rightarrow \oy_{L sj \dot \a}
\ee
\item
Then write down the result of this substitution:
\be
{\cal L}_{P,s}=p_{sq} P^q
\fr{\pa}{\pa J} + 
g   K^j \fr{\pa}{\pa L^{sj}} \Ra p_{sq} P^q
\oy_{J \dot \a} + 
g   K^j \oy_{L sj \dot \a}= \w_{P,s \dot \a} 
\ee
The result $\w_{P,s \dot \a} $ is a simple generator of a dotspinor which satisfies the constraint equation in \ci{cybersusyII}, which is the same as the constraint equation in equation (\ref{wfweffgwfw}) above:

\be
d_3 \w_{P,s \dot \a} =0
\ee

 \een
This is a consequence of the definition of the operator
\be
d_3 
= 
\ov C_{\dot \a} \fr{\pa  P_{\rm SSM} }{\pa H^i}
\ov \y^{\dag
}_{H i \dot \a}
+\cdots
\ee
which gives rise to the following mutual implications for Lie algebra generators and dotspinor generators:

\be
d_3 
 \w_{P,s \dot \a} 
=0
\Lra
{\cal L}_{P,s} P_{{\rm SSM}}   =0
\ee

\section{Collection of Basic Dotspinors}

\subsection{The five ordinary dotspinors from the five ordinary symmetry generators}

\ben
\item  Strong SU(3):
\be
 {\cal C}^A = T^{A c}_{d}\lt (
  Q^{dpi} \fr{\pa}{\pa Q^{cpi}}
+  B^{p}_{c} \fr{\pa}{\pa  B^{p}_{d} }
+   T^{p}_{c} \fr{\pa}{\pa T^{p}_{d}}\rt )
\eb
\Ra 
 T^{A c}_{d}\lt (
  Q^{dpi} \ov \y_{Qcpi\dot \a}
+  B^{p}_{c} \ov \y_{Bp\dot \a}^{d} 
+   T^{p}_{c}  \ov \y_{T p\dot \a}^{d} \rt )= \w^{A}_{{\cal C}\dot \a}
\ee
where $T^{A c}_{d}$ are the hermitian $3 \times 3$ matrix generators of the Lie Algebra of SU(3),
\item  Weak SU(2)
\be
 {\cal I}^a = t^{ai}_{j}\lt (
H^j \fr{\pa}{\pa H^i}
+K^j \fr{\pa}{\pa K^i}
+L^{pj} \fr{\pa}{\pa L^{pi}}
+Q^{cpj} \fr{\pa}{\pa Q^{cpi}}
\rt )
\eb
\Ra 
 t^{ai}_{j}\lt (
  Q^{cpj} \ov \y_{Qcpi\dot \a}
+ L^{pj} \ov \y_{Lpi\dot \a}
+ H^{j} \ov \y_{Hi\dot \a}
+ K^{j} \ov \y_{Ki\dot \a}
 \rt )= \w^{a}_{{\cal I}\dot \a}
\ee
where $t^{ai}_{j}$ are the hermitian $2 \times 2$ matrix generators of the Lie Algebra of SU(2),
\item  Hypercharge U(1)
\be
 {\cal Y} = \lt (
- H^i \fr{\pa}{\pa H^i}
+ K^i \fr{\pa}{\pa K^i}
-L^{pi} \fr{\pa}{\pa L^{pi}}
\ebp
+ 2 P^{p} \fr{\pa}{\pa P^{p}}
+ \fr{1}{3} Q^{cpi} \fr{\pa}{\pa Q^{cpi}}
+ \fr{2}{3} B^{p}_{c} \fr{\pa}{\pa  B^{p}_{c} }
- \fr{4}{3} T^{p}_{c} \fr{\pa}{\pa T^{p}_{c}}
\rt )
\eb
\Ra
 \w_{{\cal Y}\dot \a}
\ee
\item Lepton Number
\be
 {\cal L}= \lt (
L^{pi} \fr{\pa}{\pa L^{pi}}
- P^{p} \fr{\pa}{\pa P^{p}}
- R^{p} \fr{\pa}{\pa R^{p}}
\rt )
\eb
\Ra
 \w_{{\cal L}\dot \a}
\ee
\item Baryon Number
\be 
  {\cal B}=
 \fr{1}{3} Q^{cpi} \fr{\pa}{\pa Q^{cpi}}
- \fr{1}{3} B^{p}_{c} \fr{\pa}{\pa  B^{p}_{c} }
- \fr{1}{3} T^{p}_{c} \fr{\pa}{\pa T^{p}_{c}}
\eb
\Ra
 \w_{{\cal B}\dot \a}
\ee
\een

\subsection{The eight basic `J-dotspinors' from the eight basic `J-symmetry' Generators}

\subsubsection{Leptons}

\ben
\item
 The Positron Generator ${\cal L}_{P,s}$:
\be
{\cal L}_{P,s}=p_{sq} P^q
\fr{\pa}{\pa J} + 
g   K^j \fr{\pa}{\pa L^{sj}} 
\Ra p_{sq} P^q
\oy_{J \dot \a} + 
g   K^j \oy_{L sj \dot \a}= \w_{P,s \dot \a} \ee
\item
The Right Neutrino Generator ${\cal L}_{R,s}$:
\be
{\cal L}_{R,s}=r_{sq} R^q
\fr{\pa}{\pa J} + 
g   H^j \fr{\pa}{\pa L^{sj}} 
\Ra r_{sq} R^q
\oy_{J \dot \a} + 
g   H^j  \oy_{L sj \dot \a}= \w_{R,s \dot \a}\ee
\item The  Left Electron Generator ${\cal L}_{E,s}$:
\be
{\cal L}_{E,s}= p_{qs} \e_{ij} H^i L^{qj} 
\fr{\pa}{\pa J} + 
g \e_{ij} H^i K^j \fr{\pa}{\pa P^{s}} 
\eb
\Ra p_{qs} \e_{ij} H^i L^{qj} 
\oy_{J \dot \a} + 
g \e_{ij} H^i K^j \oy_{P s \dot \a}= \w_{E,s \dot \a}\ee
\item
 The  Left Neutrino Generator ${\cal L}_{N,s}$:
\be
{\cal L}_{N,s}= r_{qs} \e_{ij} K^i L^{qj} 
\fr{\pa}{\pa J} + 
g \e_{ij} H^i K^j \fr{\pa}{\pa R^{s}} 
\eb
\Ra r_{qs} \e_{ij} K^i L^{qj} 
\oy_{J \dot \a} + 
g \e_{ij} H^i K^j \oy_{R s \dot \a}= \w_{N,s \dot \a}\ee
\een
We can put these together in various ways, but that does not seem as interesting at present as putting the quarks together. 
The quarks form meson and hadrons. Combining these leptons does not seem so experimentally vital right now.

\subsubsection{Quarks}
Here are the quark dotspinors:

\ben
\item  The Bottom dotspinor:
\be
 {\cal L}_{B,sc}
=
  \lt (b_{sq} B^q_c
\fr{\pa}{\pa J} + 
g   K^j \fr{\pa}{\pa Q^{scj}} 
\rt )
\eb
\Ra
  \lt (b_{sq} B^q_c
\oy_{J \dot \a}  + 
g   K^j \oy_{Q,scj \dot \a}
\rt )
= 
\w_{B,sc \dot \a}
\ee
\item The Top dotspinor:
\be
 {\cal L}_{T,sc}
=  \lt (
t_{sq} T^q_c
\fr{\pa}{\pa J} + 
g   H^j \fr{\pa}{\pa Q^{scj}} 
\rt )
\eb
\Ra
  \lt (t_{sq} T^q_c
\oy_{J \dot \a}  + 
g   H^j \oy_{Q,scj \dot \a}
\rt )
= \w_{T,sc \dot \a}
\ee
\item The Up dotspinor:
\be 
  {\cal L}^c_{U,s}=    
t_{qs} \e_{ij} K^i Q^{cqj} 
\fr{\pa}{\pa J} + 
g \e_{ij} H^i K^j \fr{\pa}{\pa T^{s}_c} 
\eb
\Ra
  b_{qs} \e_{ij} K^i Q^{cqj} 
\oy_{J \dot \a}  + 
g \e_{ij} H^i K^j \oy_{T,s \dot \a}^c 
= \w^{c}_{U,s \dot \a}
\ee
\item The Down dotspinor:
\be
  {\cal L}^c_{D,s}=    b_{qs} \e_{ij} H^i Q^{cqj} 
\fr{\pa}{\pa J} + 
g \e_{ij} H^i K^j \fr{\pa}{\pa B^{s}_c} 
\eb
\Ra
  b_{qs} \e_{ij} H^i Q^{cqj} 
\oy_{J \dot \a}  + 
g \e_{ij} H^i K^j \oy_{B,s \dot \a}^c 
= 
\w^{c}_{D,s \dot \a}
\ee
\een
We can put these expressions $\w_{B,sc \dot \a}
, \w_{T,sc \dot \a}
, \w^{c}_{U,s \dot \a}
,\w^{c}_{D,s \dot \a}
$
together in various ways to form 
strong SU(3) invariants.  They will also be  weak SU(2) invariants. We can choose various combinations to get various values of   hypercharge ${\cal Y}$.
They have the following values of hypercharge
\be
\w_{B,sc \dot \a}: {\cal Y} = \fr{2}{3}; 
 \w_{T,sc \dot \a}: {\cal Y} = \fr{-4}{3}; 
 \w^{c}_{U,s \dot \a}: {\cal Y} = \fr{+4}{3}; 
\w^{c}_{D,s \dot \a}: {\cal Y} = \fr{-2}{3}; 
\ee
 
\section{Products of the Basic Dotspinors}
\la{fqwefewggrergeg}

A new insight arrives when we consider the 13 basic dotspinors that arise from the 13 Lie Algebra generators.  We can multiply these together to obtain new invariant dotspinors.

An element of the Lie algebra must be a sum of terms with only one derivative with respect to a field.  

However an element of the dotspinor invariants can have more than one dotspinor in a term.

So here is the correct statement:
\ben
\item
The five ordinary symmetries ${\cal C}^A,{\cal I}^a,{\cal Y},
{\cal B},{\cal L}$  generate ordinary basic dotspinors.
\item
The eight basic `J-symmetries' generate  basic J-dotspinors.
\item
The dotspinors can be multiplied by each other and by any desired functions of the scalars  to form dotspinors that still satisfy the constraints.  The multiplication must be such that the result is symmetrized over the dotted spinor indices of the result.
\item
The result can be converted to a pseudosuperfield by putting hats on everything to convert them.
\item
The components of the resulting  pseudosuperfield can be found using projections as in \ci{cybersusyII} and \ci{cybersusyIII}.
\item
Any product which contains at least one J-dotspinor is a J-dotspinor. 
\item
As was demonstrated in \ci{cybersusyIII}, any J-dotspinor  generates a cohomology tower, which implies a cybersusy algebra, which gives rise to an effective action with supersymmetry breaking. 
\item
 We shall not try to write the cohomology  towers out here, and we shall not write down the pseudosuperfields that start with the wrong spin (such as spin  $J = 1$  for a baryon, or spin $J = \fr{1}{2}$ for a meson), but we shall briefly mention some of the composites in the next sections, to illustrate how they are put together.  
\la{fqwefewggrergeg1}
\een

\section{Mesonic Combinations of the Generators}

\subsection{Leptonic Mesons}

Here is an example of a meson with ${\cal Y}=2$ and ${\cal L}={\cal B}=0$  and $J=1 $ made from the leptonic dotspinors:

\be
A_{PN,(\dot \a \dot \b)} = a_{PN}^{pq}
\lt (
\w_{P,p \dot \a}\w_{N,q \dot \b}
+
\w_{P,p \dot \b}\w_{N,q \dot \a}
\rt )
\ee

What is this?  What does it have to do with the weak vector meson $W^{+}_{J=1}$?  Its supersymmetry tower is related to the various leptonic   modes of decay for $\P^+, K^+,D^+,B^+,\cdots$ or a $\r^+,\cdots$ presumably\footnote{\large There is an important exercise to be done relating the various composites in this paper to the particle states in the Review of Particle Physics \ci{physletreview}. }.

\subsection{Hadronic Mesons}

\la{hadmesons}

The situation is more interesting when one combines the basic quark dotspinors with each other and with suitable scalars to produce operators that generate particles that are known to be hadrons.  We will choose all of the operators to be strong SU(3) invariants and weak SU(2) invariants. This discussion is introductory, and there is much more to be said. 

\ben
\item
Hadronic Mesons with ${\cal Y}=2$ and $J=1$

Consider the expression
\be
A_{BU,\dot \a \dot \b}= a_{BU}^{pq}
\lt (
\w_{B,pc \dot \a}\w^{c}_{U,q \dot \b}
+ \w_{B,pc \dot \b}\w^{c}_{U,q \dot \a}
\rt )
\la{fwefwegehgehrt}
\ee
This has the quantum numbers of a hadronic meson with spin one and hypercharge ${\cal Y}=2$ . It satisfies
\be
d_3
A_{\dot \a \dot \b}
=0
\ee
 Clearly, part of this yields zero when symmetrized:
$\oy_{J \dot \a}  \oy_{J \dot \b} + \oy_{J \dot \b}  \oy_{J \dot \a} =0 $ 
but the rest survives.  Similar remarks apply to the following:

\item
Hadronic Mesons with ${\cal Y}=-2$ and $J=1$

The following expression is a strong SU(3) invariant and a weak SU(2) invariant with ${\cal Y}=-2$:
\be
A_{TD,\dot \a \dot \b}= a_{TD}^{pq}
\lt (
\w_{T,pc \dot \a}\w^{c}_{D,q \dot \b}
+ \w_{T,pc \dot \b}\w^{c}_{D,q \dot \a}
\rt )
\ee

 \item Hadronic Mesons with ${\cal Y}=0$ and $J=1$

The following two expressions are  strong SU(3) invariants and  weak SU(2) invariants with ${\cal Y}=0$:
 \be
A_{TU,\dot \a \dot \b}= a_{TU}^{pq}
\lt (
\w_{T,pc \dot \a}\w^{c}_{U,q \dot \b}
+ \w_{T,pc \dot \b}\w^{c}_{U,q \dot \a}
\rt )
\ee
 \be
A_{BD,\dot \a \dot \b}= a_{BD}^{pq}
\lt (
\w_{B,pc \dot \a}\w^{c}_{D,q \dot \b}
+ \w_{B,pc \dot \b}\w^{c}_{D,q \dot \a}
\rt )
\ee

\een

It seems wrong to try to calculate any  masses  from these operators and their supersymmetry breaking effective actions. The main problem is that it appears that all of these mesonic operators, both leptonic and hadronic,  will mix with operators that have the same quantum numbers but are made from the gauge 
particles, so we would need the gauge theory cohomology to proceed to find the relevant complete cohomology towers that mix under cybersusy for any given set of quantum numbers.  This is why the leptons and baryons are easier to deal with.  One cannot make a lepton or a baryon using only gauge particles.

\section{Baryons}

We will now review, in a very superficial way,  some of the baryon dotspinor operators that arise from combining the quark dotspinors.
The advantage of looking at leptons and baryons is that we know that there is no combination of gauge particles that has non-zero lepton number or baryon number. So we can ignore the gauge particles in the first approximation for leptons and baryons.

That is why it was possible to calculate a mass spectrum for the leptons in Cybersusy I \ci{cybersusyI} without worrying about the gauge theory. 

It is also possible to do the same for the baryons, but it is a large task, and it has not yet been done.

All of the expressions we will look at will be strong SU(3) invariants and weak SU(2) invariants, with baryon number one or minus one, and  with various values of the hypercharge Y and various numbers of spinor indices.

\section{Supermultiplets whose lowest terms are Baryons with Spin $ J= \fr{3}{2}$}

\la{hadbaryons}

  The baryonic generators in \ci{cybersusyIII} can be obtained in this way.  

\ben
\item
The Multiplets that include $\D^{-}_{J= \fr{3}{2}}$

\ben
\item
Baryon with ${\cal Y}=-2$ and ${\cal B}=+1$

The following  expression has  ${\cal Y}=-2$, and it has baryon number ${\cal B}=1$:
 \be
\w_{DDD,\dot \a_1 }=   f^{(p_1p_2p_3)}
\ve_{c_1 c_2 c_3} 
\w^{c_1}_{D,p_1 \dot \a_1} 
\w^{c_2}_{D,p_2 \dot \a_2} 
\w^{c_3}_{D,p_3 \dot \a_3} 
\ee
This expression is automatically symmetric in the spinor indices $(\dot \a_1\dot \a_2\dot \a_3)$, so long as the coefficients $ f^{(p_1p_2p_3)}$ are chosen to be symmetric in the flavour indices $(p_1p_2p_3)$. 
\item
Baryon with ${\cal Y}=2$ and ${\cal B}=-1$

The following  expression has  ${\cal Y}=2$, and it has baryon number ${\cal B}=-1$:
 \be
\w_{BBB,(\dot \a_1\dot \a_2\dot \a_3)}=   f^{(p_1p_2p_3)}
\ve^{c_1 c_2 c_3} 
\lt (
\w_{B,p_1 c_1 \dot \a_1} 
\w_{B,p_2 c_2 \dot \a_2} 
\w_{B,p_3 c_3 \dot \a_3} 
\rt )
\ee
This expression is automatically symmetric in the spinor indices $(\dot \a_1\dot \a_2\dot \a_3)$, so long as the coefficients $ f^{(p_1p_2p_3)}$ are chosen to be symmetric in the flavour indices $(p_1p_2p_3)$. 

\een
\item
The Multiplet that includes $\D^{0}_{J= \fr{3}{2}}$

\ben

\item
Baryon with ${\cal Y}=0$ and ${\cal B}=+1$

The following  expression has  ${\cal Y}=0$, and it has baryon number ${\cal B}=1$:
 \be
\w_{DDU,(\dot \a_1\dot \a_2\dot \a_3)}=   
f^{(p_1 p_2) p_3}
\ve_{c_1 c_2 c_3} 
\eb
\lt (
 \w^{c_1}_{D,p_1 \dot \a_1} 
\w^{c_2}_{D,p_2 \dot \a_2} 
\w^{c_3}_{U,p_3 \dot \a_3} 
\ebp
+
\w^{c_1}_{D,p_1 \dot \a_3} 
\w^{c_2}_{D,p_2 \dot \a_1} 
\w^{c_3}_{U,p_3 \dot \a_2} 
\ebp
 +
\w^{c_1}_{D,p_1 \dot \a_2} 
\w^{c_2}_{D,p_2 \dot \a_3} 
\w^{c_3}_{U,p_3 \dot \a_1} 
\rt )
\ee
This expression is automatically symmetric in the spinor indices $(\dot \a_1\dot \a_2\dot \a_3)$, so long as the coefficients $ f^{(p_1p_2)p_3}$ are chosen to be symmetric in the first two flavour indices $(p_1p_2)$. 
\item
Baryon with ${\cal Y}=0$ and ${\cal B}=-1$

The following  expression has  ${\cal Y}=0$, and it has baryon number ${\cal B}=-1$:
 \be
\w_{BBT,(\dot \a_1\dot \a_2\dot \a_3)}=   
f^{(p_1 p_2) p_3} \ve^{c_1 c_2 c_3} 
\eb\lt (
\w_{B,p_1 c_1 \dot \a_1} 
\w_{B,p_2 c_2 \dot \a_2} 
\w_{T,p_3 c_3 \dot \a_3} 
\ebp
+
 \w_{B,p_1 c_1 \dot \a_2} 
\w_{B,p_2 c_2 \dot \a_3} 
\w_{T,p_3 c_3 \dot \a_1} 
\ebp
+
\w_{B,p_1 c_1 \dot \a_3} 
\w_{B,p_2 c_2 \dot \a_1} 
\w_{T,p_3 c_3 \dot \a_2} 
\rt )
\ee
This expression is automatically symmetric in the spinor indices $(\dot \a_1\dot \a_2\dot \a_3)$, so long as the coefficients $ f^{(p_1p_2)p_3}$ are chosen to be symmetric in the first two flavour indices $(p_1p_2)$. 

\een
\item
The Multiplet that includes $\D^{+}_{J= \fr{3}{2}}$
\ben

\item
Baryon with ${\cal Y}=+2$ and ${\cal B}=+1$

The following  expression has  ${\cal Y}=+2$, and it has baryon number ${\cal B}=1$:
 \be
\w_{UUD,(\dot \a_1\dot \a_2\dot \a_3)}= 
f^{(p_1 p_2) p_3} 
\ve_{c_1 c_2 c_3} 
\eb
\lt (
\w^{c_1}_{U,p_1 \dot \a_1} 
\w^{c_2}_{U,p_2 \dot \a_2} 
\w^{c_3}_{D,p_3 \dot \a_3} 
\ebp
+
\w^{c_1}_{U,p_1 \dot \a_3} 
\w^{c_2}_{U,p_2 \dot \a_1} 
\w^{c_3}_{D,p_3 \dot \a_2} 
\ebp
+
\w^{c_1}_{U,p_1 \dot \a_2} 
\w^{c_2}_{U,p_2 \dot \a_3} 
\w^{c_3}_{D,p_3 \dot \a_1} 
\rt )
\ee
This expression is automatically symmetric in the spinor indices $(\dot \a_1\dot \a_2\dot \a_3)$, so long as the coefficients $ f^{(p_1p_2)p_3}$ are chosen to be symmetric in the first two flavour indices $(p_1p_2)$. 

\item
Baryon with ${\cal Y}=-2$ and ${\cal B}=-1$

The following  expression has  ${\cal Y}=-2$, and it has baryon number ${\cal B}=-1$:
 \be
\w_{TTB,(\dot \a_1\dot \a_2\dot \a_3)}= f^{(p_1 p_2) p_3} 
\ve^{c_1 c_2 c_3} 
\lt (
\w_{T,p_1 c_1 \dot \a_1} 
\w_{T,p_2 c_2 \dot \a_2} 
\w_{B,p_3 c_3 \dot \a_3} 
\ebp
+
\w_{T,p_1 c_1 \dot \a_2} 
\w_{T,p_2 c_2 \dot \a_3} 
\w_{B,p_3 c_3 \dot \a_1} 
+\w_{T,p_1 c_1 \dot \a_3} 
\w_{T,p_2 c_2 \dot \a_1} 
\w_{B,p_3 c_3 \dot \a_2} 
\rt )
\ee
This expression is automatically symmetric in the spinor indices $(\dot \a_1\dot \a_2\dot \a_3)$, so long as the coefficients $ f^{(p_1p_2)p_3}$ are chosen to be symmetric in the first two flavour indices $(p_1p_2)$. 
\een

\item
The Multiplets that include $\D^{++}_{J= \fr{3}{2}}$
\ben
\item
Baryon with ${\cal Y}=+4$ and ${\cal B}=+1$

The following  expression has  ${\cal Y}=+4$, and it has baryon number ${\cal B}=1$:
 \be
\w_{UUU,(\dot \a_1\dot \a_2\dot \a_3)}=   f^{(p_1 p_2 p_3)} 
\ve^{c_1 c_2 c_3} 
\w^{c_1}_{U,p_1 \dot \a_1} 
\w^{c_2}_{U,p_2 \dot \a_2} 
\w^{c_3}_{U,p_3 \dot \a_3} 
\ee
This expression is automatically symmetric in the spinor indices $(\dot \a_1\dot \a_2\dot \a_3)$, so long as the coefficients $ f^{(p_1p_2p_3)}$ are chosen to be symmetric in the flavour indices $(p_1p_2p_3)$. 
\item
Baryon with ${\cal Y}=-4$ and ${\cal B}=-1$

The following  expression has  ${\cal Y}=-4$, and it has baryon number ${\cal B}=-1$:
 \be
\w_{TTT,(\dot \a_1\dot \a_2\dot \a_3)}=   f^{(p_1 p_2 p_3)} 
\ve^{c_1 c_2 c_3} 
\lt (
\w_{T,p_1 c_1 \dot \a_1} 
\w_{T,p_2 c_2 \dot \a_2} 
\w_{T,p_3 c_3 \dot \a_3} 
\rt )
\ee
This expression is automatically symmetric in the spinor indices $(\dot \a_1\dot \a_2\dot \a_3)$, so long as the coefficients $ f^{(p_1p_2p_3)}$ are chosen to be symmetric in the flavour indices $(p_1p_2p_3)$. 
\een
\een

\section{Supermultiplets whose lowest terms are Baryons with Spin $ J= \fr{1}{2}$}

\la{hadhalfbaryons}

\ben

\item
{Some Multiplets that include $\Xi^-_{b, J= \fr{1}{2}}$: the singlet made of $dsb$ quarks with $J= \fr{1}{2}$ }
\ben
\item
{Baryon with ${\cal Y}=-2$ and ${\cal B}=+1$}

The following  expression has  ${\cal Y}=-2$, and it has baryon number ${\cal B}=1$:
 \be
\w_{DDD,\dot \a}=   f^{[p_1p_2]p_3}
\e_{c_1 c_2 c_3} 
\eb
(\ve_{i_1 j_1} Q^{c_1 i_1}_{p_1} H^{j_1} )
(\ve_{i_2 j_2} Q^{c_2 i_2}_{p_2} H^{j_2} )
\w^{c_3}_{D,p_3 \dot \a} 
\ee
Note that the coefficients $f^{[p_1p_2]p_3}$ can be chosen to be antisymmetric in the indices $[p_1p_2]$, because the expression itself is antisymmetric in the indices $[p_1p_2]$, due to the contraction with the antisymmetric tensor $\e_{c_1 c_2 c_3} $.

\item{Baryon with ${\cal Y}=2$ and ${\cal B}=-1$}

The following  expression has  ${\cal Y}=+2$, and it has baryon number ${\cal B}=-1$:
 \be
\w_{BBB,\dot \a}=  f^{[p_1p_2]p_3}
\ve^{c_1 c_2 c_3} 
 B^{p_1}_{c_1} 
 B^{p_2}_{c_2} 
\w_{B,p_3 c_3 \dot \a} 
\ee

Note that the coefficients $f^{[p_1p_2]p_3}$ can be chosen to be antisymmetric in the indices $[p_1p_2]$, because the expression itself is antisymmetric in the indices $[p_1p_2]$, due to the contraction with the antisymmetric tensor $\e_{c_1 c_2 c_3} $.

\een

\item
{The Multiplets that include the neutron $N^0_{ J= \fr{1}{2}}$}
\ben
\item
{A First Baryon with ${\cal Y}=0$ and ${\cal B}=+1$}

The following  expression has  ${\cal Y}=0$, and it has baryon number ${\cal B}=1$:
 \be
\w_{DDU,\dot \a }=   f^{[p_1p_2]p_3}
\ve_{c_1 c_2 c_3} 
(\ve_{i_1 j_1} Q^{c_1 i_1}_{p_1} H^{j_1} )
(\ve_{i_2 j_2} Q^{c_2 i_2}_{p_2} H^{j_2} )
\w^{c_3}_{U,p_3 \dot \a} 
\ee
Note that the coefficients $f^{[p_1p_2]p_3}$ in the above can be chosen to be antisymmetric in the indices $[p_1p_2]$, because the expression itself is antisymmetric in the indices $[p_1p_2]$, due to the contraction with the antisymmetric tensor $\e_{c_1 c_2 c_3} $.

\item
{A Second Baryon with ${\cal Y}=0$ and ${\cal B}=1$}

The following  expression also has  ${\cal Y}=0$, and it has baryon number ${\cal B}=1$:
 \be
\w_{DUD,\dot \a }=   f^{p_1p_2p_3}
\ve_{c_1 c_2 c_3} 
\eb
(\ve_{i_1 j_1} Q^{c_1 i_1}_{p_1} H^{j_1} )
(\ve_{i_2 j_2} Q^{c_2 i_2}_{p_2} K^{j_2} )
\w^{c_3}_{D,p_3 \dot \a} 
\ee
Note that the coefficients $f^{p_1p_2p_3}$ in the above have no symmetry properties, because the three expressions that are contracted with the antisymmetric tensor $\e_{c_1 c_2 c_3} $ are all different.

\item
{A First Baryon with ${\cal Y}=0$ and ${\cal B}=-1$}

The following  expression has  ${\cal Y}=0$, and it has baryon number ${\cal B}=-1$:
 \be
\w_{BBT,\dot \a}=   f^{[p_1p_2]p_3}
\ve^{c_1 c_2 c_3} 
\lt (
 B^{p_1}_{c_1} 
 B^{p_2}_{c_2} 
\w_{T,p_3 c_3 \dot \a} 
\rt )
\ee
Note that the coefficients $f^{[p_1p_2]p_3}$ can be chosen to be antisymmetric in the indices $[p_1p_2]$, because the expression itself is antisymmetric in the indices $[p_1p_2]$, due to the contraction with the antisymmetric tensor $\e_{c_1 c_2 c_3} $.
\item
{A Second Baryon with ${\cal Y}=0$ and ${\cal B}=-1$}

The following  expression also has  ${\cal Y}=0$, and it has baryon number ${\cal B}=-1$:
 \be
\w_{BTB,\dot \a}=  f^{p_1p_2p_3}
\ve^{c_1 c_2 c_3} 
\lt (
 B^{p_1}_{c_1} 
 T^{p_2}_{c_2} 
\w_{B,p_3 c_3 \dot \a} 
\rt )
\ee
Note that the coefficients $f^{p_1p_2p_3}$ in the above have no symmetry properties, because the three expressions that are contracted with the antisymmetric tensor $\e_{c_1 c_2 c_3} $ are all different.

\een

\item
{The Multiplets that include the proton $P^{+}_{ J= \fr{1}{2}}$}
\ben
\item
{A First Baryon with ${\cal Y}=+2$ and ${\cal B}=+1$}

The following  expression has  ${\cal Y}=+2$, and it has baryon number ${\cal B}=1$:
 \be
\w_{UUD,\dot \a}=   f^{[p_1p_2]p_3}
\ve_{c_1 c_2 c_3} 
(\ve_{i_1 j_1} Q^{c_1 i_1}_{p_1} K^{j_1} )
(\ve_{i_2 j_2} Q^{c_2 i_2}_{p_2} K^{j_2} )
\w^{c_3}_{D,p_3 \dot \a} 
\ee
Note that the coefficients $f^{[p_1p_2]p_3}$ can be chosen to be antisymmetric in the indices $[p_1p_2]$, because the expression itself is antisymmetric in the indices $[p_1p_2]$, due to the contraction with the antisymmetric tensor $\e_{c_1 c_2 c_3} $.

\item
{A Second Baryon with ${\cal Y}=+2$ and ${\cal B}=+1$}

The following  expression also has  ${\cal Y}=+2$, and it has baryon number ${\cal B}=1$:
 \be
\w_{UDU,\dot \a}=   f^{p_1p_2p_3}
\ve_{c_1 c_2 c_3} 
(\ve_{i_1 j_1} Q^{c_1 i_1}_{p_1} K^{j_1} )
(\ve_{i_2 j_2} Q^{c_2 i_2}_{p_2} H^{j_2} )
\w^{c_3}_{D,p_3 \dot \a} 
\ee

Note that the coefficients $f^{p_1p_2p_3}$ in the above have no symmetry properties, because the three expressions that are contracted with the antisymmetric tensor $\e_{c_1 c_2 c_3} $ are all different.

\item
{A First Baryon with ${\cal Y}=-2$ and ${\cal B}=-1$}

The following  expression has  ${\cal Y}=-2$, and it has baryon number ${\cal B}=-1$:
 \be
\w_{TTB,\dot \a}=   f^{[p_1p_2]p_3}
\ve^{c_1 c_2 c_3} 
 T^{p_1}_{c_1} 
 T^{p_2}_{c_2} 
\w_{B,p_3 c_3 \dot \a} 
\ee
Note that the coefficients $f^{[p_1p_2]p_3}$ can be chosen to be antisymmetric in the indices $[p_1p_2]$, because the expression itself is antisymmetric in the indices $[p_1p_2]$, due to the contraction with the antisymmetric tensor $\e_{c_1 c_2 c_3} $.

\item
{A Second Baryon with ${\cal Y}=-2$ and ${\cal B}=-1$}

The following  expression also has  ${\cal Y}=-2$, and it has baryon number ${\cal B}=-1$:
 \be
\w_{TBT,\dot \a}=  f^{p_1p_2p_3}
\ve^{c_1 c_2 c_3} 
\lt (
 T^{p_1}_{c_1} 
 B^{p_2}_{c_2} 
\w_{T,p_3 c_3 \dot \a} 
\rt )
\ee
\een
Note that the coefficients $f^{p_1p_2p_3}$ in the above have no symmetry properties, because the three expressions that are contracted with the antisymmetric tensor $\e_{c_1 c_2 c_3} $ are all different.

\item
The Multiplets that include $\Xi^{++}_{t, J= \fr{1}{2}}$: the singlet made of $tcu$ quarks with $J= \fr{1}{2}$ 
\ben

\item{Baryon with ${\cal Y}=+4$ and ${\cal B}=+1$}

The following  expression has  ${\cal Y}=+4$, and it has baryon number ${\cal B}=1$:
 \be
\w_{UUU,(\dot \a}=   f^{[p_1p_2]p_3}
\ve_{c_1 c_2 c_3} 
\eb
(\ve_{i_1 j_1} Q^{c_1 i_1}_{p_1} K^{j_1} )
(\ve_{i_2 j_2} Q^{c_2 i_2}_{p_2} K^{j_2} )
\w^{c_3}_{U,p_3 \dot \a} 
\ee
Note that the coefficients $f^{[p_1p_2]p_3}$ can be chosen to be antisymmetric in the indices $[p_1p_2]$, because the expression itself is antisymmetric in the indices $[p_1p_2]$, due to the contraction with the antisymmetric tensor $\e_{c_1 c_2 c_3} $.

\item{Baryon with ${\cal Y}=-4$ and ${\cal B}=-1$}

The following  expression has  ${\cal Y}=-4$, and it has baryon number ${\cal B}=-1$:
 \be
\w_{TTT,\dot \a}=   f^{[p_1p_2]p_3}
\ve^{c_1 c_2 c_3} 
\lt (
 T^{p_1}_{c_1} 
 T^{p_2}_{c_2} 
\w_{T,p_3 c_3 \dot \a} 
\rt )
\ee
Note that the coefficients $f^{[p_1p_2]p_3}$ can be chosen to be antisymmetric in the indices $[p_1p_2]$, because the expression itself is antisymmetric in the indices $[p_1p_2]$, due to the contraction with the antisymmetric tensor $\e_{c_1 c_2 c_3} $.

\een

\een

These operators, together with bosonic operators and the cybersusy algebras  that arise from them, will be discussed further in the next paper of this series.

\section{Discussion of Spontaneous Breaking of Gauge Symmetry}

\la{fqerghjiortbhbrt}

Next we want to rederive the algebra used in \ci{cybersusyI} to generate supersymmetry breaking.  We will introduce a different notation here from what was used there.  This notation is intermediate between the unbroken theory and the broken theory and so is a little easier to use than either of those for present purposes.

The term $- g_{\rm J} m^2 J$ 
 gives rise to gauge symmetry breaking if $g_{\rm J}\neq 0$. We note that
\be
\ov F'_J = \fr{\pa \; P_{{\rm SP}}  }{\pa J} = 
\ve_{ij} 
g  H^i K^j 
- g_{\rm J} m^2 
\ee
needs a shift of the scalar field parts of the superfield to eliminate the $m^2$ term:
\be
H^1 \ra (mv + H^1)
, 
K^2 \ra (mv + K^2)
\la{substtt}
\ee
Here is our new notation. Let us write this in the form:
\be
H^i \ra (mv h^i + H^i)
, 
K^i \ra (mv k^i + K^i)
\ee
where
\be
h^1=1,h^2=0,k^1=0,k^2=1
\ee

Then we have a zero vacuum expectation value for the auxiliary field:
\be
<\ov G_J>_{VEV}   = \ve_{ij} 
g < H^i K^j >_{VEV}
- g_{\rm J} m^2 \eb
= 
m^2 v^2  \ve_{ij}  h^i k^j 
- g_{\rm J} m^2 = 
g m^2 v^2 ( h^1 k^2 - h^2 k^1 )
- g_{\rm J} m^2 =0
\ee
which means that supersymmetry is conserved by this VEV.

This is the development of a vacuum expectation value (VEV) of $m v$ in these two fields, followed by a shift to fields with no vacuum expection value.  Here we have 
\be
v^2 = \fr{g_{\rm J}}{g}
\la{tgqergreojirge}
\ee 

We need to rewrite the action after the substitution 
in equation (\ref{substtt}), in terms of new eigenstates of mass and charge.

The  superpotential, after this shift and redefinition, is:
\be
P_{\rm SSM}   =
g \e_{ij}(m v h^i + H^i) (m v k^j + K^j) J
- g_{\rm J} m^2 J
\eb
+
p_{p q} \e_{ij} L^{p i} (m v h^j + H^j)  P^{ q} 
+
r_{p q} \e_{ij} L^{p i}(m v k^j + K^j) R^{ q}
\eb
+
t_{p  q} \e_{ij} Q^{c p i} (m v k^j + K^j) T_c^{ q}
+
b_{p  q} \e_{ij} Q^{c p i} (m v h^j + H^j) B_c^{ q}
\la{suppotentialshifted}
\ee

In the alternative, we can define  mass and charge eigenstates, in terms of the shifted fields, and write:
\be
K = \fr{1}{\sqrt{2}}
\lt ( H^1 - K^2
\rt )
\ee+
\be
H = \fr{1}{\sqrt{2}}
\lt ( H^1 + K^2
\rt )
\ee
This has the inverse
\be
K^2 = \fr{1}{\sqrt{2}}\lt (H-K\rt )
\ee
\be
H^1 = \fr{1}{\sqrt{2}}\lt (H+K\rt )
\ee
In terms of the original action, however, this amounts to the substitution:
\be
K^2 \ra m v +  \fr{1}{\sqrt{2}}\lt (H-K\rt )
\ee
\be
H^1 \ra mv +  \fr{1}{\sqrt{2}}\lt (H+K\rt )
\ee

and  written in terms of charge eigenstates it is:
\[
P_{\rm SSM } =
\]
\[
  \lt ( g  m v  \sqrt{2} \; H +   g   \fr{1}{2} ( H H - K K ) -g  H^- K^+ \rt )J
\]
\[
+
p_{pq} \lt [ N^{p} H^- -E^{p} \lt \{ mv + \fr{1}{\sqrt{2}}\lt (H+K\rt )\rt \} \rt ]
 P^{q}
\]
\[
+
r_{pq} \lt [ N^{p} \lt \{ mv + \fr{1}{\sqrt{2}}\lt ( H-K\rt )
\rt \} -E^{p} K^+ \rt ]
 {R}^{q} 
\]
\[
+
t_{pq} \lt [ U^{c p } \lt \{ mv + \fr{1}{\sqrt{2}}\lt (H-K\rt )
\rt \} -D^{c p } K^+ \rt ] 
T_c^q
\]
\be
+
b_{pq} \lt [ U^{c p } H^- -D^{c p } \lt \{ mv + \fr{1}{\sqrt{2}}\lt (H+K\rt )\rt \} \rt ] 
 B_c^q
\la{gqeghrtrtwhrth}
\ee

In the above $H$ and $J$ constitute a massive Higgs boson supermultiplet, and $K,K^{+},H^{-}$ are three Goldstone Boson supermultiplets which will be `eaten' by the vector boson supermultiplets to form the massive weak vector boson supermultiplets $Z^{0}, W^{+},W^{-}$.

\section{Typical Example of generation of the cybersusy algebra}

When we perform the shifts in the operators and in the superpotential, we find that the superpotential is no longer invariant under the action of the cybersusy generators. For example consider
\be
{\cal L}_{P,s}=p_{sq} P^q
\fr{\pa}{\pa J} + 
g  (mvk^j+K^j)  \fr{\pa}{\pa L^{sj}} 
\ee
Then we get:
\[
{\cal L}_{P,s}P_{\rm SSMB} =
\lt \{
p_{sq} P^q
\fr{\pa}{\pa J} + 
g   (mvk^j+K^j) \fr{\pa}{\pa L^{sj}} \rt \}
\]
\[
\lt \{
g \e_{ij}
(mvh^i+H^i)K^j J
+g \e_{ij}
( H^i)(mvk^j+K^j) J
\rt.
\]\[
+
p_{p q} \e_{ij} L^{p i} (mvh^j+H^j)  P^{ q} 
+
r_{p q} \e_{ij} L^{p i} (mvk^j+K^j)R^{ q}
\]\be 
\lt.
+
t_{p  q} \e_{ij} Q^{c p i} (mvk^j+K^j) T_c^{ q}
+
b_{p  q} \e_{ij} Q^{c p i} (mvh^j+H^j) B_c^{ q}
\rt \}
\ee
This is

\[
{\cal L}_{P,s}P_{\rm SSMB} =
g \e_{ij}
(mvh^i+H^i)K^j p_{sq} P^q
+g \e_{ij}
( H^i)(mvk^j+K^j) p_{sq} P^q
\]
\[
+
p_{s q} \e_{ij} g   (mvk^i+K^i)   (mvh^j+H^j)  P^{ q} 
+
r_{s q} \e_{ij} g   (mvk^i+K^i)   (mvk^j+K^j)R^{ q}
\]
\[
=
g \e_{ij}
(mvh^i+H^i)K^j p_{sq} P^q
+g \e_{ij}
( H^i)(mvk^j+K^j) p_{sq} P^q
\]
\[
+
p_{s q} \e_{ij} g   (mvk^i )   (mvh^j )  P^{ q} 
\]
\be
+
p_{s q} \e_{ij} g   ( K^i)   (mvh^j+H^j)  P^{ q} 
+
p_{s q} \e_{ij} g   (mvk^i+K^i)   ( H^j)  P^{ q} 
\ee
\be
=
p_{s q} \e_{ij} g   (mvk^i )   (mvh^j )  P^{ q} 
=
-p_{s q}   g   m^2 v^2  P^{ q} 
=
- m^2 g_J p_{s q}       P^{ q} 
\ee
In summary, expressing things in terms of the shifted fields makes no difference.  The result is that
\be
{\cal L}_{P,s}P_{\rm SSMB} =
- m^2 g_J p_{s q}       P^{ q} 
\ee
and we get the same thing more easily from 
\be
{\cal L}_{P,s}P_{\rm SSMB}  =\lt \{
p_{sq} P^q
\fr{\pa}{\pa J} + 
g   K^j  \fr{\pa}{\pa L^{sj}} \rt \}
\lt \{ P_{\rm SSM}  - g_J m^2 J\rt \}
\eb
=
p_{sq} P^q
\fr{\pa}{\pa J}  
\lt \{ - g_J m^2 J\rt \}
=- m^2 g_J p_{s q}       P^{ q} 
\ee
\section{List of Dotspinor Cybersusy Algebra for each basic J-Invariance after Gauge Symmetry Breaking}

Using the relations
\be
{\cal L}_{P,s}=p_{sq} P^q
\fr{\pa}{\pa J} + 
g   K^j \fr{\pa}{\pa L^{sj}} 
\Ra p_{sq} P^q
\oy_{J \dot \a} + 
g   K^j \oy_{L sj \dot \a}= \w_{P,s \dot \a} \ee
we see that this means that
\be
d_3  \w_{P,s \dot \a}  =
d_3
\lt \{ p_{sq} P^q
\oy_{J \dot \a} + 
g   K^j \oy_{L sj \dot \a}
\rt \}
=- m^2 g_J p_{s q}       P^{ q} 
\ov C_{\dot \a}
\ee
and this in turn means that for the pseudosuperfields, we have:
\be
\d  {\hat \w}_{P,s \dot \a}  
=- m^2 g_J p_{s q}       {\hat P}^{ q} 
\ov C_{\dot \a}
\ee

All eight of the cybersusy generators behave in the same way. Here is a complete list:

\be
{\cal L}_{P,s}P_{\rm SSMB} =
- m^2 g_J p_{s q}       P^{ q} 
\Lra 
d_3  \w_{P,s \dot \a}  
=- m^2 g_J p_{s q}       P^{ q} 
\ov C_{\dot \a}
\ee
\be
{\cal L}_{R,s}P_{\rm SSMB} =
- m^2 g_J r_{s q}       R^{ q} 
\Lra 
d_3  \w_{R,s \dot \a}  
=- m^2 g_J r_{s q}       R^{ q} 
\ov C_{\dot \a}\ee
\be
{\cal L}_{N,s}P_{\rm SSMB} =
- m^2 g_J r_{qs}       {\cal N}^{ q} 
\Lra 
d_3  \w_{N,s \dot \a}  
=- m^2 g_J r_{qs}       {\cal N}^{ q} 
\ov C_{\dot \a}
\ee
\be
{\cal L}_{E,s}P_{\rm SSMB} =
- m^2 g_J p_{qs}       {\cal E}^{ q} 
\Lra 
d_3  \w_{E,s \dot \a}  
=- m^2 g_J p_{qs}       {\cal E}^{ q} 
\ov C_{\dot \a}
\ee

\be
{\cal L}_{T,s c}P_{\rm SSMB} =
- m^2 g_J t_{sq}       T^{ q}_c 
\Lra 
d_3  \w_{T,s,c \dot \a}  
=- m^2 g_J t_{s q}       T^{ q}_c 
\ov C_{\dot \a}
\ee
\be
{\cal L}_{B,s c}P_{\rm SSMB} =
- m^2 g_J b_{sq}       B^{ q}_{c} 
\Lra 
d_3  \w_{B,s,c \dot \a}  
=- m^2 g_J b_{s q}       B^{ q}_c 
\ov C_{\dot \a}\ee
\be
{\cal L}_{U,s}^{c}P_{\rm SSMB} =
- m^2 g_J t_{qs}       {\cal U}^{ qc} 
\Lra 
d_3  \w_{U,s,c \dot \a}  
=- m^2 g_J t_{qs}      {\cal U}^{ qc} 
\ov C_{\dot \a}
\ee
\be
{\cal L}_{D,s}^cP_{\rm SSMB} =
- m^2 g_J b_{qs}       {\cal D}^{ qc} 
\Lra 
d_3  \w_{D,s,c \dot \a}  
=- m^2 g_J b_{qs}      {\cal D}^{ qc} 
\ov C_{\dot \a}
\ee

where we define the composite scalars:
\be
{\cal N}^p= \e_{ij}(m v k^i + K^i )L^{jp}
\ee
\be
{\cal E}^p= \e_{ij}(m v h^i + H^i )L^{jp}
\ee
\be
{\cal U}^{pc}= \e_{ij}(m v k^i + K^i )Q^{jcp}
\ee
\be
{\cal D}^{pc}= \e_{ij}(m v h^i + H^i )Q^{jcp}
\ee

\section{Typical Example of continued invariance of the non-cybersusy generators}

The invariance generator $I^a$ does not have a $\fr{\pa}{\pa J}$ term in it, so it remains an invariance of the potential even after the term $ - g_J m^2 J$ is added to ${\cal L}_{P,s}P_{\rm SSM}$.

\be
{\cal L}_{P,s}P_{\rm SSMB}  =
 P_{\rm SSM}  - g_J m^2 J
\ee

Here are the collected and rearranged
results, written in terms of the shifted fields:
\be
I^a
{\cal L}_{P,s}P_{\rm SSMB}
\eb
=
g \e_{ij}  t^{ai}_{k} (mvh^k +  H^k)
(mv k^j +  K^j) J
\eb
+g \e_{ij}(mvh^i +  H^i) 
t^{aj}_{k}(mv k^k +  K^k) J
\eb
+
p_{p q} \e_{ij} L^{p i}  
t^{aj}_{k} (mv h^k +  H^k)
 P^{ q} 
\eb
+
p_{p q} \e_{ij} t^{ai}_{k}L^{pk}  (mvh^j +  H^j) P^{ q} 
\eb
+
r_{p q} \e_{ij} L^{p i}  
t^{aj}_{k}(mv k^k +  K^k) R^{ q}
\eb
+
r_{p q} \e_{ij} t^{ai}_{k}L^{pk}  (mv k^j +  K^j) R^{ q}
\eb
+
t_{p  q} \e_{ij} Q^{c p i} 
t^{aj}_{k}(mv k^k +  K^k) T_c^{ q}
\eb
+
t_{p  q} \e_{ij} t^{ai}_{k}Q^{cpk}  (mv k^j +  K^j) T_c^{ q}
\eb
+
b_{p  q} \e_{ij} t^{ai}_{k}Q^{cpk}  (mvh^j +  H^j) B_c^{ q}
\eb
+
b_{p  q} \e_{ij} Q^{c p i}    
t^{aj}_{k} (mv h^k +  H^k) B_c^{ q}
=0
\ee
The continued invariance follows from the invariance properties of the invariant tensor $\e_{ij}$ under SU(2).  The shift does nothing, because it comes from the term $ - g_J m^2 J$, and there is no $\fr{\pa}{\pa J}$ term in the five well-known invariances of the standard model.

\section{The cybersusy algebras and their effective actions}
We note that:
\ben
\item
The ordinary symmetries of the standard model, such as $SU(3) \times SU(2) \times U(1)$, lepton number and baryon number, do  give rise to dotspinor solutions to the constraint equations, but those solutions do not give rise to the cybersusy algebra, and so they do not give rise to supersymmetry breaking as discussed in 
the four papers \ci{cybersusyI}\ci{cybersusyII}\ci{cybersusyIII}\ci{cybersusyIV}.
\item
This means that, for  particles that are not generated by the basic generators in subsection  \ref{adfafdsfljjklasdf}, supersymmetry breaking  must be sought in the cohomology of supersymmetric gauge theory.  That is a question for the future.
\item
We have seen that the ordinary generators reviewed above {\bf do not lead to the cybersusy algebra}.  Generators that have a term $\fr{\pa}{\pa J}$ in them {\bf do
 lead to the cybersusy algebra}. 
\item
All of the above composite mesons and baryons and the composite leptons too will generate towers of cohomology analagous to the towers seen in \ci{cybersusyIII}.  It is relatively straightforward to generate the cybersusy algebra and effective actions from any of them.  There are a number of different such cybersusy algebras.  Each of them requires a paper of its own to analyze the situation, and there will be challenges too, since the flavours will complicate things, particularly for mixed charge states like the neutron and proton.  
\item
Note that for the mesons there is a chance that the familiar mesons will be the lowest mass particles, in accord with experimental results, since these towers begin with spin one. The baryon towers begin with spin $\fr{3}{2}$ and the lepton tower analyzed in  \ci{cybersusyI} and \ci{cybersusyIV} began with spin $\fr{1}{2}$.  So these are all very different from each other, and each will require quite a lot of work.  
 \een

\section{Other Composite Operators}

Suppose that one has got an operator that can create a 
given supermultiplet containing a particle of interest, say the operator in equation (\ref{fwefwegehgehrt}), for example.  Then it is evident that there are an infinite number of other operators that will create the same particle.  One can obtain them by multiplying the operator (\ref{fwefwegehgehrt}) by any possible collection of scalar fields that does not change its quantum numbers.  And it is obvious that such an operation will not change the cybersusy algebra associated to that operator, since only the dotspinors generate the cybersusy algebra, and we are assuming that we do not change those.

So in some sense it appears that the cybersusy algebra affects all the operators which create a given supermultiplet, and hence it is a real phenomenon determining the supersymmetry breaking for that supermultiplet. There are of course other operators that arise from the cohomology when we include derivatives, as discussed in \ci{cybersusyII} and  \ci{cybersusyIII}.  It seems a reasonable conjecture that these do not change the situation in respect of this general phenomenon.

\section{A supersymmetric type of Quark Model}

We have seen that we can generate mesons and hadrons using the fields $\w_{B,sc \dot \a}, \w_{T,sc \dot \a}
, \w^{c}_{U,s \dot \a}
$ and $\w^{c}_{D,s \dot \a}$ just as though they were quarks, except that they must appear with their Lorentz spinor indices symmetrized.  When it would be appropriate to contract Lorentz spinor indices, for example to make spin $J= \fr{1}{2}$
baryons, we use the appropriate scalar fields $B_{s}^{c}$ or 
$T_{s}^{c}$ or ${\cal D}^{cp} =\e_{ij} H^i Q^{jcp} $ or 
${\cal U}^{cp} =\e_{ij}K^i Q^{jcp} $
instead.  But the whole thing closely follows the quark model, except that it is being done with the first terms of chiral superfields instead of quarks. 

The rest of cybersusy then follows the superfields.  The useful  development is that the BRS cohomology results in effective supersymmetry anomalies in the effective theory that are generated by gauge symmetry breaking.  From there we can calculate the masses after supersymmetry breaking, as was done in \ci{cybersusyIV} for the leptons.

\section{Back to soft supersymmetry breaking terms?}

\la{erjiopejioperge}

Cybersusy generates a kind of supersymmetric version of the quark model, together with a supersymmmetric version of the leptons.

Because of the existence of the dotspinor multiplets for the leptons and the quarks, it seems fairly clear that one can manufacture anything out of these dotspinors that one could expect to find in the quark model with leptons in the standard model.

And because of the basic result for the leptons in the first four papers on cybersusy \ci{cybersusyI}\ci{cybersusyII}\ci{cybersusyIII}\ci{cybersusyIV}, it is also clear that in some sense it is natural for the quarks and the leptons to be by far the lightest particles in their supermultiplets. Exactly the same algebra applies to the quarks as was found for the leptons, except that the matrices change.  For the quarks there are colour indices as well as flavour indices that are extant, but the colour indices are not coupled to the supersymmetry breaking--they just ride along free.

It follows that one could apply cybersusy directly to the quarks. Since they have exactly the same cybersusy algebra as the leptons, we already know exactly what that will yield from the analysis of the leptons.  The quarks will naturally be the lightest particles in their supermultiplets, just as the leptons were.  And the difference in mass from the next lightest supersymmetry partner can be as large as we want it to be, at least at first glance.

 Now it is true that when the quarks are combined into hadrons or mesons with the colour indices contracted to make colour singlets, one generates a new type of cybersusy algebra, and that corresponds to a different effective action, and the spectrum of that probably looks quite different from what one would obtain by simply replacing the squark masses with new soft supersymmetry breaking masses.   

It seems very likely that the new effective action will 
generate a different spectrum than one would obtain from the SSM with soft supersymmetry breaking by squark masses. 

But, in a sense, what we have found, is a way of justifying the method of introducing soft supersymmetry breaking terms for the sleptons and the squarks. This is the same method that we criticised in section \ref{fewfwfwefwefkthklhrtkl} of this paper.  Now we have found it again, except that there are more superpartners than appear in the SSM, and for some regimes these might make a difference.  It also seems likely that looking at the effect of cybersusy on various mesons and baryons might generate quite a different spectrum from what one would get by constructing them out of massive squarks etc. So that work still needs to be done.

What is clearly completely missing from cybersusy, so far, is an explanation of the supersymmetry breaking for the massive weak vector bosons, the photon, and the Higgs. There is also a puzzle relating to calculations in QCD, which  have been extensive \ci{ellis}.  Should the gauginos for the color gauge theory be taken to be massive for some reason?  Should one include massless superpartners in these calculations?  
Perhaps some, or all, of these questions can be understood better if one looks at the BRS cohomology of gauge theories. 

 \section{Conclusion}

Cybersusy may give rise to a consistent theory of supersymmetry breaking, in time, but there is much to be done before that can be decided.  The only positive result, so far, is that the electrons and neutrinos, as shown in the first four papers on cybersusy \ci{cybersusyI}\ci{cybersusyII}\ci{cybersusyIII}\ci{cybersusyIV}, and the quarks, as shown in this paper, do seem to naturally be the lightest particles in their broken supermultiplets.  Moreover, the splitting between these elementary leptons and quarks, and the lightest superpartners for them, can be huge. 

So far, there are no negative results for cybersusy. However it is disappointing that there are still so many parameters in the theory.  Moreover cybersusy gives rise to lots of new particles. It is true that if one could observe them all, there would be enough correlations to verify cybersusy, or to contradict it, but that would take a lot of data. Also it is clear that there is no explanation yet of the huge and tiny numbers  needed to get a reasonable scale for supersymmetry breaking.  

It is also interesting that the standard supersymmetric model seems to be set up in a way that allows cybersusy to act nicely on the quarks and leptons, through the quark and lepton symmetries that we found in subsection \ref{adfafdsfljjklasdf} of this paper.   It is not obvious whether there are other models that would work just as well, and that is a question worth examination.

 In order for the program to continue to be successful, it is also necessary that the BRS cohomology of gauge theories be understood better, and that it has more cybersusy algebra associated with it in some way.

 It is encouraging that a sort of cybersusy quark model is emerging, because that means that there is a reasonable chance to construct all the baryons in a way that manifests broken supersymmetry. There are many calculations to be done in this baryon area, and it would be nice to know if these are consistent with experiment.  The program for this is briefly recalled  in section \ref{fqwefewggrergeg} above.  The first step is to write down the cohomology towers, then derive the cybersusy algebra, then write out the effective action, and then invert the kinetic terms to find the propagators and the masses. A first glance at this is given in sections 
\ref{hadmesons},
\ref{hadbaryons} and
\ref{hadhalfbaryons} above.

 That is a {\bf real test of the theory, because it could well happen that these effective actions give results that are obviously wrong}--for example one might find that the theory predicts that the mesons are predicted to be heavier than their spin $J=\fr{1}{2}$ superpartners, or that the baryons are predicted to be heavier than their spin  $J=0$ baryon superpartners. That would be a real problem for cybersusy, and maybe also for supersymmetry itself, in terms of its application to particle physics.  The data that need to be correlated are already immense and very well documented\ci{physletreview}.

\np

\tableofcontents

\end{document}